\documentclass[pra,twocolumn,showpacs,preprintnumbers,amsmath,amssymb]{revtex4}
\usepackage{graphicx}
\usepackage{dcolumn}
\usepackage{bm}
\usepackage{amsthm}
\usepackage{color}
\input xy
\xyoption{all}

\newcommand{\ket}[1]{\left| #1 \right\rangle}
\newcommand{\bra}[1]{\left\langle #1 \right |}

\newcommand{\g}{\mathfrak{g}}

\newtheorem{theorem}{Theorem}

\newtheorem{proposition}{Proposition}

\newtheorem{corollary}{Corollary}
\DeclareMathOperator{\tr}{tr}

\DeclareMathOperator{\Stab}{Stab}

\newcommand{\R}{{\mathbb R}}

\newcommand{\Id}{{\rm Id}}

\begin{document}

\title{Classification of nonproduct states with maximum stabilizer dimension}

\author{David W. Lyons}
  \email{lyons@lvc.edu}
\author{Scott N. Walck}
  \email{walck@lvc.edu}
\author{Stephanie A. Blanda}
  \email{sab002@lvc.edu}
\affiliation{Lebanon Valley College, Annville, PA 17003}

\date{11 September 2007}

\begin{abstract}
Nonproduct $n$-qubit pure states with maximum dimensional stabilizer
subgroups of the group of local unitary transformations are precisely
the generalized $n$-qubit Greenberger-Horne-Zeilinger states and their
local unitary equivalents, for $n\geq 3, n\neq 4$.  We characterize the
Lie algebra of the stabilizer subgroup for these states.  For $n=4$,
there is an additional maximal stabilizer subalgebra, not local unitary
equivalent to the former.  We give a canonical form for states with this
stabilizer as well.
\end{abstract}


\maketitle

\section{Introduction}

The desire to measure and classify entanglement for multiparty states of
$n$-qubit systems has been motivated by potential applications in
quantum computation and communication that utilize entanglement as a
resource~\cite{nielsenchuang,gudder03}.  Entanglement classification is also
an interesting question in the fundamental theory of quantum
information.  

It is natural to consider two states to have the same entanglement type
if one is transformable to the other by a local unitary
transformation~\cite{linden98, linden99}. Thus any entanglement measure,
as a function on state space, must be invariant under the action of the
local unitary group.  One such invariant is the isomorphism class of the
stabilizer subgroup (the set of local unitary transformations that do
not alter a given state) and its Lie algebra of infinitesimal
transformations.  Inspired by ideas originally laid out for 3-qubit
systems~\cite{linden98, carteret00a}, the authors of the present paper
have achieved a number of results for systems of arbitrary numbers of
qubits on the structure of stabilizer Lie subalgebras and in particular,
those stabilizers that have maximum possible dimension.  Evidence that
maximum stabilizer dimension is an interesting property is the fact that
states that have such stabilizers also turn out to maximize other known
entanglement measures and play key roles in quantum computational
algorithms~\cite{higuchi00, brierley07}. In~\cite{irredent} we describe a
connection between stabilizer structure and the question of when a pure
state is determined by its reduced density matrices.

States with maximum stabilizer dimension are products of singlet pairs
(with an unentangled qubit when the number of qubits is odd) and their
local unitary equivalents~\cite{minorb1, minorb2}.  In the
present paper, we show an analogous result for {\em nonproduct} states
with maximum stabilizer dimension.  These are the generalized $n$-qubit
Greenberger-Horne-Zeilinger (GHZ) states
$$\alpha\ket{00\cdots 0} + \beta\ket{11\cdots 1}, \hspace{.2in}
\alpha,\beta\neq 0$$
and their local unitary equivalents for 3 or more qubits, together
with an additional class of states for the special case of 4 qubits.
In order to prove these results, we extend the stabilizer analysis of our
previous work~\cite{maxstabnonprod1} from stabilizers of pure states to
stabilizers of arbitrary density matrices. 

\section{The main result}

The $n$-qubit local unitary group is 
$$G = G_0 \times G_1\times \cdot \times G_n$$
where $G_0 = U(1)$ is the group of phases and $G_j = SU(2)$ is the
3-dimensional group
of single qubit rotations in qubit $j$ for $1\leq j\leq n$.  Its Lie
algebra 
$$\g = \g_0 \oplus \g_1 \oplus \cdots \oplus \g_n$$
is the set of infinitesimal local unitary transformations acting on the
tangent level,
where $\g_0$ is 1-dimensional and each $\g_j=su(2)$, the set of
skew-Hermitian matrices with trace zero.  Given an $n$-qubit state
vector $\ket{\psi}$, let 
$$\Stab_\psi=\{g\in G\colon g \ket{\psi} = \ket{\psi}\}$$ 
denote the subgroup of elements in $G$
that stabilize $\psi$, and let 
$$K_\psi = \{X\in \g \colon X \ket{\psi} = 0\}$$ be the corresponding Lie
algebra.

In~\cite{maxstabnonprod1} we
show that the maximum possible dimension for $K_\psi$ is $n-1$ when
$\psi$ is not a product state, for $n\geq 3$.  Further, we show that for
$n\neq 4$, we must have $\dim P_j K_\psi = 1$ for all $j$, where
$P_j\colon \g \to \g_j$ is the natural projection.  In fact this
condition holds for the generalized $n$-qubit GHZ state
$$\alpha \ket{00\cdots 0} + \beta\ket{11\cdots 1}$$ and its LU
equivalents.  For $n=4$, there is an additional possible
condition for $K_\psi$ to attain maximum dimension, and that is $\dim
P_j K_\psi = 3$ for all $j$ and $K_\psi\cong su(2)$.  The main result of
the present paper is the converse of these previous results.  We show
that the generalized $n$-qubit GHZ states (and LU equivalents) are the
{\em only} nonproduct states that have maximum stabilizer dimension,
except for the 4-qubit case.  For those 4-qubit states with stabilizer
isomorphic to $su(2)$, we give a canonical representative for each LU
equivalence class.  Here is the formal statement.

\begin{theorem}\label{mainthm}
{\bf (Classification of states with maximal stabilizer)} Let
  $\ket{\psi}$ be a nonproduct $n$-qubit state with stabilizer subalgebra
  $K_\psi$ of maximum possible dimension $n-1$, for some $n\geq 3$.  One
  of two conditions must hold:
  \begin{enumerate}
\item    [(i)]  $\dim P_j K_\psi = 1$ for $1\leq j\leq n$, or
\item [(ii)] $n=4$, $\dim P_j K_\psi = 3$ for $1\leq j\leq n$
and $K_\psi\cong su(2)$.
  \end{enumerate}
If~(i) holds, then $\ket{\psi}$ is LU equivalent to a generalized
$n$-qubit GHZ state.  If~(ii) holds, then $\ket{\psi}$ is LU
equivalent to a state of the form
\begin{eqnarray*}
\ket{\psi}&=& a(\ket{0011}+\ket{1100})\\
& +& b(\ket{1001} + \ket{0110})\\
& +& c(\ket{1010} + \ket{0101})
\end{eqnarray*}
for some complex coefficients $a,b,c$ satisfying $a> 0$, $abc\neq 0$ and
$a+b+c=0$.  Furthermore, this is a unique representative of the LU
equivalence class for $\ket{\psi}$ of this form.
\end{theorem}

The proof divides naturally into two sections, one for each of the two
conditions on stabilizer in the statement of the theorem.  We consider
conditions~(i) and~(ii) in sections~\ref{1dim} and~\ref{4qubit},
respectively.  It is convenient to work in terms of density matrices and
their local unitary stabilizers; the following preliminary section
establishes the necessary extension of our previous results to local
unitary action on density matrices.

\section{Preliminary propositions on stabilizer structure}\label{stabstructsection}

For the set of $n$-qubit pure and mixed density matrices, we may omit
the phase factor in local unitary operations, and take the
local unitary group to be 
$$SU(2)^n = G_1\times \cdots \times G_n \subset G$$ 
and its Lie algebra to be
$$su(2)^n = \g_1\oplus \cdots \oplus \g_n\subset \g.$$
Given an $n$-qubit density matrix $\rho$, an element $g\in
SU(2)^n$ acts on $\rho$ by
$$g\cdot \rho = g \rho g^\dag.$$
An element $X\in su(2)^n$ acts on the infinitesimal level by
$$X\cdot \rho = [X,\rho] = X\rho - \rho X.$$
Let 
$$\Stab_\rho=\{g\in SU(2)^n\colon g \rho g^\dag = \rho\}$$ 
denote the subgroup of elements
that stabilize $\rho$, and let 
$$K_\rho = \{X\in su(2)^n \colon [X ,\rho] = 0\}$$ be the corresponding
Lie algebra.  We use multi-index notation $I=(i_1,i_2,\ldots,i_n)$,
where each $i_k$ is a binary digit, to denote labels for the standard
computational basis for state space.  We write $I^c$ to denote the
bitwise complement of $I$.

One expects a natural correspondence between $K_\psi$ and $K_\rho$ in
the case where $\rho=\ket{\psi}\bra{\psi}$, and indeed there is.  

\begin{proposition}\label{kpsiprojtokrho}
  Let $\rho = \ket{\psi}\bra{\psi}$ be a pure $n$-qubit density matrix.
  Then $K_\rho = P K_\psi$, where $P\colon \g \to \g_1\oplus \cdots
  \oplus \g_n$ is the natural projection that
  drops the phase factor.
\end{proposition}

\begin{proof}
We begin with a simple observation.  Let $X\in \g$ and let
$\ket{\phi}=X\ket{\psi}$.  Skew-Hermicity of $X$ gives us
\begin{equation}
  \label{miniobservation}
\bra{\psi}X = -\bra{\phi}.
\end{equation}

To see that $PK_\psi\subset K_\rho$, let $X\in K_\psi$, say
$X=(-it,X_1,X_2,\ldots, X_n)$.  Let $Y=PX$, so
$Y\ket{\psi}=it\ket{\psi}$.  Then
$$[Y,\rho]=Y\ket{\psi}\bra{\psi} - \ket{\psi}\bra{\psi}Y = it\rho
-it\rho = 0.$$ 
To get the second equality,
apply~(\ref{miniobservation}) to $\ket{\phi}=it\ket{\psi}=Y\ket{\psi}$.

Conversely,
we show that $ K_\rho\subset PK_\psi$.  Let $Y\in K_\rho$, and let
$\ket{\phi}=Y\ket{\psi}$.  Using~(\ref{miniobservation}) again, we have
\begin{eqnarray*}
0&=&[Y,\rho]=Y\ket{\psi}\bra{\psi} - \ket{\psi}\bra{\psi}Y\\
 &=&
\ket{\phi}\bra{\psi} + \ket{\psi}\bra{\phi} = \ket{\phi}\bra{\psi} +
(\ket{\phi}\bra{\psi})^\dag
\end{eqnarray*}
so $\ket{\phi}\bra{\psi}$ is skew-Hermitian.  Thus there is some unitary
$U$ such that $U\ket{\phi}\bra{\psi}U^\dag$ is diagonal, with pure
imaginary entries.  In fact there is only one nonzero diagonal entry, say $is$ for
some real $s$, since $\ket{\phi}\bra{\psi}$ has rank one.  Therefore
$U\ket{\phi} = is U\ket{\psi}$, so $\ket{\phi}=is\ket{\psi}$.
Thus $Y$ is the projection of $(-is,Y_1,Y_2,\ldots,Y_n)$ in $K_\psi$.
This concludes the proof.
\end{proof}

Let us use the symbol $P_j$
to denote both the projection $P_j\colon \g \to \g_j$ and also $P_j\colon
su(2)^n\to \g_j$.  Applying $P_j$ to both sides of $K_\rho = PK_\psi$
yields the following corollary.

\begin{corollary}
  \label{pjkrhoispjkpsi}
  Let $\rho = \ket{\psi}\bra{\psi}$ be a pure $n$-qubit density matrix.
  Then $P_j K_\rho = P_j K_\psi$ for $1\leq j\leq n$.
\end{corollary}

The next proposition establishes that we may work with either $K_\rho$
or $K_\psi$ to calculate stabilizer dimension.  

\begin{proposition}\label{dimstabssame}
  Let $\rho = \ket{\psi}\bra{\psi}$ be a pure $n$-qubit density matrix.
Then $\dim K_\rho = \dim K_\psi$.
\end{proposition}

\begin{proof}
Let $X_1,\ldots,X_r \in K_\psi$ be linearly independent.  Let
$Y_k=PX_k$ for $1\leq k\leq r$ and suppose that 
$$0 = \sum a_k Y_k$$
for some scalars $a_k$.  Since $Y_k = X_k - it_k$ for some $t_k$, we have
$$\sum a_kX_k = i \sum a_kt_k.$$
Since $\sum a_kX_k$ is in $K_\psi$, so is $i \sum a_kt_k$, and so both
of these sums must be zero.  Since the $X_k$ are independent, all the
$a_k$ must be zero, and hence the $Y_k$ are also linearly independent.
\end{proof}

We choose
the basis 
\begin{align*}
A &= \left[ \begin{array}{cc} i & 0 \\ 0 & -i \end{array} \right] , &
B &= \left[ \begin{array}{cc} 0 & 1 \\ -1 & 0 \end{array} \right] , &
C &= \left[ \begin{array}{cc} 0 & i \\  i & 0 \end{array} \right] .
\end{align*}
for $su(2)$.  We denote by $A_j$ the element in $\g$ that has
$A$ in the $j$th qubit slot and zeros elsewhere.  By slight abuse of
notation, we use the same symbol for the corresponding element in
$su(2)^n$.  Analogous notation applies for $B_j$ and $C_j$, for $1\leq
j\leq n$.

We will make repeated use of the following basic calculation.  Let
$X=\sum t_kA_k$ be an element of the local unitary Lie algebra
$su(2)^n$.  It is straightforward to check that
\begin{equation}
  \label{basicactioncalc}
X\cdot\rho = [X,\rho] = \sum_{I,J}\zeta(I,J)\rho_{I,J}\ket{I}\bra{J},
\end{equation}
where
$$\zeta(I,J) = i\sum_{\ell =1}^nt_\ell [(-1)^{i_\ell} +
  (-1)^{j_{\ell}+1}] = 2i \sum_{\ell\colon i_\ell \neq j_\ell}
  (-1)^{i_\ell}t_\ell.
$$

As a consequence, we have the following.

\begin{proposition}\label{asactforce}
Let $\rho$ be an $n$-qubit density matrix, let $X=\sum t_kA_k$, and
suppose that $X\in K_\rho$.  Then either $\zeta(I,J)=0$ or
$\rho_{I,J}=0$ for all $I,J$.  
\end{proposition}

We record here a proposition proved in~\cite{minorb2} (Lemma~3.8) with
an alternative proof using Proposition~\ref{asactforce}.

\begin{proposition}\label{unentqubitcrit}
  {\bf (Stabilizer criterion for an unentangled
  qubit)}
Let $\rho=\ket{\psi}\bra{\psi}$ be a pure $n$-qubit density matrix where
$\ket{\psi}=\sum c_I\ket{I}$.  If $A_\ell\in K_\rho$, then the $\ell$th
  qubit is unentangled.
\end{proposition}

\begin{proof}
  Choose any nonzero state coefficient $c_{I'}$.  Apply
  Proposition~\ref{asactforce} to $X=A_\ell$, $I=I'$ and $J$ for which
  $j_\ell \neq i'_\ell$.  Since $\zeta(I',J)=\pm 2i\neq 0$, we conclude
  that $c_J$ must be zero.
\end{proof}

\begin{proposition}\label{nqubitghzstab}
  {\bf (Stabilizer criterion for generalized $n$-qubit GHZ states)} Let
  $\ket{\psi}=\sum_I c_I\ket{I}$ be
a nonproduct $n$-qubit state vector for some $n\geq 3$, and let
$$V= \left\{\sum_{k=1}^n t_kA_k \colon \sum t_k = 0\right\}.$$
We have $V=K_\psi$ if and only if $\psi$ is a generalized $n$-qubit GHZ
state. 
\end{proposition}

\begin{proof}
In Section~V of~\cite{maxstabnonprod1} we show that if $\psi$ is a
generalized $n$-qubit GHZ state, then $K_\psi=V$.  Now we prove the
converse.

Suppose that $K_\psi=V$.  Choose a multi-index $I$ such that $c_I\neq
0$.  If $J$ is another multi-index different from $I$ in indices ${\cal
  K}=\{k_1,\ldots,k_p\}$ and equal to $I$ in some index $r$, let 
$$X = \sum_{j\in {\cal K}}(-1)^{i_j} A_j - \left(\sum_{j\in {\cal
      K}}(-1)^{i_j}\right)A_r,$$ so that the sum of the coefficients of
      the $A_j$ is zero, so $X$ is in $K_\psi$.  Viewing $X$ as an
      element of $K_\rho$ and applying~(\ref{basicactioncalc}), we have
$$\zeta(I,J) = 2i \sum_{\ell\colon i_\ell \neq j_\ell}
  (-1)^{i_\ell}t_\ell = 2i \sum_{j=1}^p (-1)^{2i_{k_j}} = 2ip \neq 0$$
so $\rho_{I,J}=c_Ic_J^\ast = 0$ by Proposition~\ref{asactforce}, so $c_J=0$.  It
  follows that $\ket{\psi}$ has the form $\ket{\psi}= \alpha \ket{I} +
  \beta\ket{I^c}$ for some nonzero $\alpha,\beta$.  Finally, if
  $I$ does not consist of all zeros or all ones, say $i_k=0, i_\ell=1$,
  then let $X=A_k-A_\ell$, then $\zeta(I,I^c) = 4i\neq 0$, so
  $\rho_{I,I^c}$ would have to be zero, but this is impossible.
  Thus we conclude that $\ket{\psi}= \alpha\ket{00\cdots 0} +
  \beta\ket{11\cdots 1}$.
\end{proof}

\section{Maximal stabilizers with 1-dimensional projections in each qubit}\label{1dim}

In this section we consider the consequences of condition~(i) in the
statement of Theorem~\ref{mainthm}.  By virtue of
Propositions~\ref{pjkrhoispjkpsi} and~\ref{dimstabssame}, we may work
with $K_\rho$ in place of $K_\psi$.  If $\dim P_j K_\rho = 1$ for all
$j$, we can apply an LU transformation so that $P_j X = t(X)A$ for all
$X\in K_\rho$, where $t(X)$ is a scalar that depends on $X$.  After this
adjustment, the stabilizer $K_\rho$ is a codimension 1 subspace of the
subspace of $su(2)^n$ spanned by $A_1,A_2,\ldots,A_n$.  Thus there is
some nonzero real vector $(m_1,m_2,\ldots, m_n)$ such that
$$K_\rho = \left\{\sum_{k=1}^n t_kA_k \colon \sum m_kt_k=0\right\}.$$ 
In the following Proposition, we show that the only way for a nonproduct
state to have this stabilizer is to be (LU equivalent to) a generalized
$n$-qubit GHZ state.

\begin{proposition}
\label{goodkernel} 
Let $\rho=\ket{\psi}\bra{\psi}$ be a pure nonproduct
$n$-qubit density matrix with stabilizer subalgebra
$$K_\rho =\left\{\sum_{k=1}^n t_kA_k \colon \sum m_kt_k=0\right\}$$ where
$0\neq(m_1,m_2,\ldots,m_n)\in \R^n$.  Then all the $m_j$ have the same absolute value and
$\ket{\psi}$ is LU equivalent to a generalized $n$-qubit GHZ state.
\end{proposition}

\begin{proof}
We may suppose, after a suitable LU transformation, that $c_{00\cdots
0}\neq 0$ (the LU transformation $\exp(\pi/2 C_j)=\Id\otimes \cdots
\otimes \Id\otimes C_j \otimes \Id \otimes \cdots \otimes \Id $ sends
$\ket{I}$ to $\ket{I_j}$, where $\Id$ is the $2\times 2$ identity matrix
and $I_j$ denotes the multi-index obtained from $I$ by complementing the
$j$th index, and changes the stabilizer element $\sum t_k A_k$ to $\sum
(-1)^{\delta_{jk}}t_k A_k$, where $\delta_{jk}$ denotes the Kronecker
delta).  This operation may change the sign of some of the $m_j$, but
does not change their absolute values.

If any $m_\ell=0$, then $A_\ell\in K_\rho$.  By
Proposition~\ref{unentqubitcrit}, it follows that the $\ell$th qubit is
unentangled.  But $\ket{\psi}$ is not a product, so we can rule out the
possibility that any $m_\ell$ is zero.

Suppose now that there exist two coordinates of
$(m_1,\ldots,m_n)$ with different absolute values.  Then
$(m_1,\ldots,m_n)$ and $J=(j_1,\ldots,j_n)$ are linearly independent,
where $J$ is any nonzero vector whose entries are all 0's and 1's.  Let
$J$ be a multi-index not equal to $00\cdots 0$ with 1's in positions
$k_1,\ldots,k_p$.  We may choose a vector $(t_1,\ldots,t_n)$ that is
perpendicular to $(m_1,\ldots,m_n)$ but {\em not} perpendicular to $J$.
It follows that $X=\sum t_kA_k$ lies in $K_\rho$, but $\sum_{\ell = 1}^p
t_{k_\ell} \neq 0$, so we conclude that $c_J= 0$ by
Proposition~\ref{asactforce}.  This means that $\ket{\psi}$ is the
completely unentangled state $\ket{00\cdots 0}$.  But this is a
contradiction, since $\ket{\psi}$ is not a product.
This establishes that all the $m_j$ have the same absolute
value.  

Thus $\ket{\psi}$ has (possibly after an LU transformation) the
stabilizer of a generalized $n$-qubit GHZ state, and therefore
$\ket{\psi}$ must be a generalized $n$-qubit GHZ state by
Proposition~\ref{nqubitghzstab}.
\end{proof}

\section{The 4-qubit case}\label{4qubit}

In this section we consider condition~(ii) of Theorem~\ref{mainthm}.  It
follows from Lemma~1 and Proposition~1 of~\cite{maxstabnonprod1} that
after an LU transformation, if necessary, we may take $K_\rho$ to be
$$K_\rho=\left\langle \sum_{k=1}^4 A_k, \sum_{k=1}^4
B_k,\sum_{k=1}^4C_k\right\rangle.$$  From this assumption, we derive a
canonical form for unique representatives of each LU equivalence class.

\begin{proposition}\label{4qubitprop}
{\bf (Classification of 4-qubit states with $su(2)$ stabilizer)}
Let $\rho=\ket{\psi}\bra{\psi}$ be a pure 4-qubit
density matrix, where $\ket{\psi}=\sum_I c_I\ket{I}$, and let 
$$V=\left\langle \sum_{k=1}^4 A_k, \sum_{k=1}^4
B_k,\sum_{k=1}^4C_k\right\rangle.$$  Then $K_{\rho}=V$ if and only if
\begin{eqnarray*}
\ket{\psi}&=& a(\ket{0011}+\ket{1100})\\
& +& b(\ket{1001} + \ket{0110})\\
& +& c(\ket{1010} + \ket{0101})
\end{eqnarray*}
for some complex coefficients
$a,b,c$ satisfying $a> 0$, $abc\neq 0$ and $a+b+c=0$.  Furthermore, this is
a unique representative of the LU equivalence class for $\ket{\psi}$ of
this form.
\end{proposition}

\begin{proof}
Suppose that $K_\rho = V$.  

First we consider the consequences of the element $\sum A_k$ in
$K_\rho$.  By Proposition~\ref{asactforce}, for each
$I,J$ for which $\rho_{I,J}=c_Ic_J^\ast\neq 0$, we have
$$\zeta(I,J) = \sum_{\ell\colon
i_\ell \neq j_\ell} (-1)^{i_\ell} = 0.$$
From this it follows that if
$c_I,c_J\neq 0$ and we have to flip $n_0$ zeros and $n_1$ ones to transform
$I$ into $J$, then $n_0=n_1$.  If $m$ is the number of indices which are
zeros in both $I$ and $J$, then the number of zeros in $I$ is $n_0+m$,
and the number of zeros in $J$ is $n_1+m = n_0+m$.  It follows that the number
of zeros is constant for all multi-indices $I$ for which $c_I\neq 0$.

Next we consider $\sum C_k$ in $K_\rho$.  Since $\exp(\pi/2 C)=C$, we 
have
$$\rho = \exp (\pi/2\sum C_k)\cdot \rho =
\sum_{I,J}\rho_{I^c,J^c}\ket{I}\bra{J},$$
so 
\begin{equation}
  \label{ceffect}
c_Ic_J^\ast = c_{I^c}c_{J^c}^\ast
\end{equation}
for all $I,J$.  Applying~(\ref{ceffect}) to
the pair $I,I$, we obtain $|c_I|=|c_{I^c}|$.
Applying~(\ref{ceffect}) to the pair $I,I^c$, we see that
$c_Ic_{I^c}^\ast$ is real, so it must be that $c_I = c_{I^c}$ for all
$I$.  Since the number of zeros is constant for
all multi-indices for nonzero state coefficients, it follows that this
number must be two. Thus we have so far that $\ket{\psi}$ must be
of the form
\begin{eqnarray*}
\ket{\psi}&=& a(\ket{0011}+\ket{1100})\\
& +& b(\ket{1001} + \ket{0110})\\
& +& c(\ket{1010} + \ket{0101})
\end{eqnarray*}
for some $a,b,c$.  If $a=0$, then $A_1-A_2$ would be in $K_\rho$,
violating our assumption.  Similarly, neither $b$ nor $c$ can be zero.

To see that $a+b+c$ must equal zero, note that $\left(\sum C_k\right)
\cdot \rho =0$ means that $\left(\sum C_k\right) \cdot \ket{\psi} = it
\ket{\psi}$ for some $t$.  A straightforward calculation shows that
$$\left(\sum C_k\right)\cdot \ket{\psi} = i \sum_I \left(\sum_k
c_{I_k}\right)\ket{I},$$
where $I_k$ denotes the multi-index obtained from $I$ by complementing
the $k$th index.  The coefficient of $\ket{1000}$ on the right hand side is
$(a+b+c)\ket{1000}$, but zero on the left hand side, so $a+b+c=0$.
Finally, we may take $a$ to be positive by applying a phase adjustment,
if necessary.

Conversely, suppose that $\psi$ has the form
\begin{eqnarray*}
\ket{\psi}&=& a(\ket{0011}+\ket{1100})\\
& +& b(\ket{1001} + \ket{0110})\\
& +& c(\ket{1010} + \ket{0101})
\end{eqnarray*}
for some $a,b,c$, $abc\neq 0$ and $a+b+c=0$.  It is easy to check that
$K_\rho$ contains $V$.  By our analysis of stabilizer structure
in~\cite{maxstabnonprod1}, we have $\dim K_\rho\leq 6$.  We know $\dim
K_\rho=6$ if and only if $\ket{\psi}$ is (LU equivalent to) a product of
two singlet pairs~(\cite{minorb2}), but we can rule this out (it is easy
to check that $\ket{\psi}$ is not a product).  Lemma~1
in~\cite{maxstabnonprod1} rules out the possibility that $\dim K_\rho$
could be 4 or 5, so $\dim K_\rho$  must equal 3 and we must have $K_\rho
= V$.

To show that $\ket{\psi}$ is the unique representative of its LU class
with the form stated in the Proposition, we use local unitary
invariants.  First, we consider invariants of the form $\tr((\tr_A
\rho)^2)$ where $A$ is a subsystem consisting of a subset of the 4
qubits.  From these trace invariants, a straightforward derivation
produces invariants $I_1,I_2,I_3$ in the table below.
\begin{center}
\begin{tabular}{c|c}
Qubit pair in $A$ & Invariant\\ \hline
$(1,2)$ & $I_{1}(\psi) = \vert a \vert \vert b \vert$ \\
$(1,3)$ & $I_{2}(\psi) = \vert a \vert \vert c \vert$ \\
$(1,4)$ & $I_{3}(\psi) = \vert b \vert \vert c \vert$ \\
\end{tabular}
\end{center}
It follows that the norms of the state coefficients for $\ket{\psi}$ are
LU invariant (for example, we have $|a|=\sqrt{I_1(\psi) I_2(\psi)/ I_3(\psi)}$).
Since we are taking $a$ to be positive, and since
$a+b+c=0$, a simple geometric or algebraic argument shows that $b$ and $c$ are
determined up to conjugate.  

To determine the imaginary parts of $b$ and
$c$, we use a class of invariants given by
$$P^m_{\sigma,\tau,\phi}(\psi) = \sum_{I^1,I^2,\ldots,I^m} c_{I^1}\cdots
c_{I^m} c^\ast_{J^1}\cdots c^\ast_{J^m}$$ where $I^1,\ldots, I^m$ is an
$m$-tuple of 4-qubit multi-indices $I^k=(i^k_1,\ldots,i^k_4)$, and $$J^k
= (i^k_1,i^k_{\sigma(1)},i^k_{\tau(1)},i^k_{\phi(1)})$$
for $1\leq k\leq m$~\cite{linden98, gingrich02}.
We consider the invariant for $m=3, \sigma = \left( \begin{tabular}{ccc}
1 & 2 & 3 \\ 
3 & 2 & 1 \\ 
\end{tabular}\right)  , \tau = \left( \begin{tabular}{ccc}
1 & 2 & 3 \\ 
2 & 1 & 3 \\ 
\end{tabular}\right)$, and $\varphi = \left( \begin{tabular}{ccc}
1 & 2 & 3 \\ 
2 & 3 & 1 \\ 
\end{tabular}\right)$.
The imaginary part of this invariant acting on $\psi$ is
$$-24a^2b_1b_{2}(b_{1}^{2} +b_{2}^{2}+ab_1)$$
where $b_1,b_2$ are the real and imaginary parts of $b$, respectively.
From this we see that if $b_1\neq 0$, then $b_2$ is determined.  If
$b_1=0$, then $b_2$ is not completely determined; there remains the
ambiguity of conjugation for $b$ and $c$.  To settle this, let
$\alpha(r,s,t)$ denote the state
\begin{eqnarray*}
& &r(\ket{0011}+\ket{1100})\\
& +& s(\ket{1001} + \ket{0110})\\
& +& t(\ket{1010} + \ket{0101})
\end{eqnarray*}
for state coefficients $r,s,t$.  The (not local unitary) operation that
permutes qubits 3 and 4 takes $\alpha(a,b,c)$ to $\alpha(a,c,b)$.  If
$b$ is pure imaginary, then $c$ is not, so our invariant
$P^3_{\sigma,\tau,\phi}$ tells us that $\alpha(a,c,b)$ is {\em not} LU
equivalent to $\alpha(a,c^\ast,b^\ast)$.  It follows that
$\alpha(a,b,c)$ is {\em not} LU equivalent to $\alpha(a,b^\ast,c^\ast)$,
for if the LU operation $g_1\otimes g_2\otimes g_3\otimes g_4$ takes
$\alpha(a,b,c)$ to $\alpha(a,b^\ast,c^\ast)$, then the LU operation
$g_1\otimes g_2\otimes g_4\otimes g_3$ takes $\alpha(a,c,b)$ to the
LU inequivalent state $\alpha(a,c^\ast,b^\ast)$. 

Having exhausted all cases, we see that $\ket{\psi}$ is the unique state in its LU
equivalence class of the form stated in the Proposition.
\end{proof}

\section{Conclusion}

We have described the stabilizer subalgebra structure and have a
complete description of states with stabilizers of maximum dimension,
both for product and nonproduct states.  We have shown evidence that
warrants further study of stabilizer structure.  There are at least
three natural directions in which to further pursue this analysis: seek
complete descriptions, perhaps a classification, of all possible
stabilizer subalgebras; seek complete description and perhaps canonical
LU forms in the spirit of Proposition~\ref{4qubitprop} for states that
have those stabilizers; and extend these pursuits to mixed states.

\section{Acknowledgments}

The authors thank the National Science Foundation
for their support of this work through NSF Award No. PHY-0555506.



\end{document}